\documentclass{article}

\usepackage{microtype}
\usepackage{graphicx}
\usepackage{subcaption}
\usepackage{booktabs} 

\usepackage{hyperref}

\usepackage[preprint]{icml2026}

\usepackage{amsmath}
\usepackage{amssymb}
\usepackage{mathtools}
\usepackage{amsthm}

\usepackage[capitalize,noabbrev]{cleveref}

\theoremstyle{plain}

\theoremstyle{definition}

\theoremstyle{remark}

\usepackage[textsize=tiny]{todonotes}

\usepackage{multirow}
\usepackage{xspace}

\newcommand{\ours}{\textsc{Speech-Omni-Lite}\xspace}

\icmltitlerunning{Speech-Omni-Lite: Portable Speech Interfaces for Vision–Language Models}

\begin{document}

\twocolumn[
  \icmltitle{\ours: Portable Speech Interfaces for Vision–Language Models}



  \icmlsetsymbol{equal}{*}

  \begin{icmlauthorlist}
    \icmlauthor{Dehua Tao}{hw}
    \icmlauthor{Xuan Luo}{polyu,hitsz}
    \icmlauthor{Daxin Tan}{hw}
    \icmlauthor{Kai Chen}{hkust}
    
    \icmlauthor{Lanqing Hong}{hw}
    \icmlauthor{Jing Li}{polyu}
    \icmlauthor{Ruifeng Xu}{hitsz,loop}
    \icmlauthor{Xiao Chen}{hw}

  \end{icmlauthorlist}

  \icmlaffiliation{hw}{Huawei Leibniz Research Center, Hong Kong SAR}
  \icmlaffiliation{polyu}{The Hong Kong Polytechnic University, Hong Kong SAR, China}
  \icmlaffiliation{hitsz}{Harbin Institute of Technology, Shenzhen, China}
  \icmlaffiliation{hkust}{Hong Kong University of Science and Technology, Hong Kong SAR, China}
  \icmlaffiliation{loop}{Shenzhen Loop Area Institute, Shenzhen, China}

  \icmlcorrespondingauthor{Xiao Chen}{chen.xiao2@huawei.com}

  \icmlkeywords{multimodal, omni-models, discrete speech token, spoken dialogue}

  \vskip 0.3in
]



\printAffiliationsAndNotice{}  

\begin{abstract}
    While large-scale omni-models have demonstrated impressive capabilities across various modalities, their strong performance heavily relies on massive multimodal data and incurs substantial computational costs. This work introduces \textbf{\ours}, a cost-efficient framework for extending pre-trained Visual-Language (VL) backbones with speech understanding and generation capabilities, while fully preserving the backbones' vision-language performance. Specifically, the VL backbone is equipped with two lightweight, trainable plug-and-play modules, a speech projector and a speech token generator, while keeping the VL backbone fully frozen. To mitigate the scarcity of spoken QA corpora, a low-cost data construction strategy is proposed to generate Question–Text Answer–Text–Speech (QTATS) data from existing ASR speech-text pairs, facilitating effective speech generation training. Experimental results show that, even with only thousands of hours of speech training data, \ours achieves excellent spoken QA performance, which is comparable to omni-models trained on millions of hours of speech data. Furthermore, the learned speech modules exhibit strong transferability across VL backbones.
    

\end{abstract}

\section{Introduction}

In the era of large-scale models, there has been growing interest in multimodal models that unify text, image, speech, and video within a single foundation model, termed \textbf{omni-model}. While omni-models~\cite{gpt-4o, Baichuan, Qwen3-omni} demonstrate excellent performance across modalities, their success typically relies on large-scale training, which demands massive multimodal datasets
and substantial computational resources. This poses a practical barrier for most research groups and motivates a resource-conscious question: \textit{\textbf{how can we extend a well-trained foundation model to new modalities in a cost-efficient manner, minimizing both data acquisition and computational expense?
}}

\begin{figure}
    \centering
    \includegraphics[width=\linewidth]{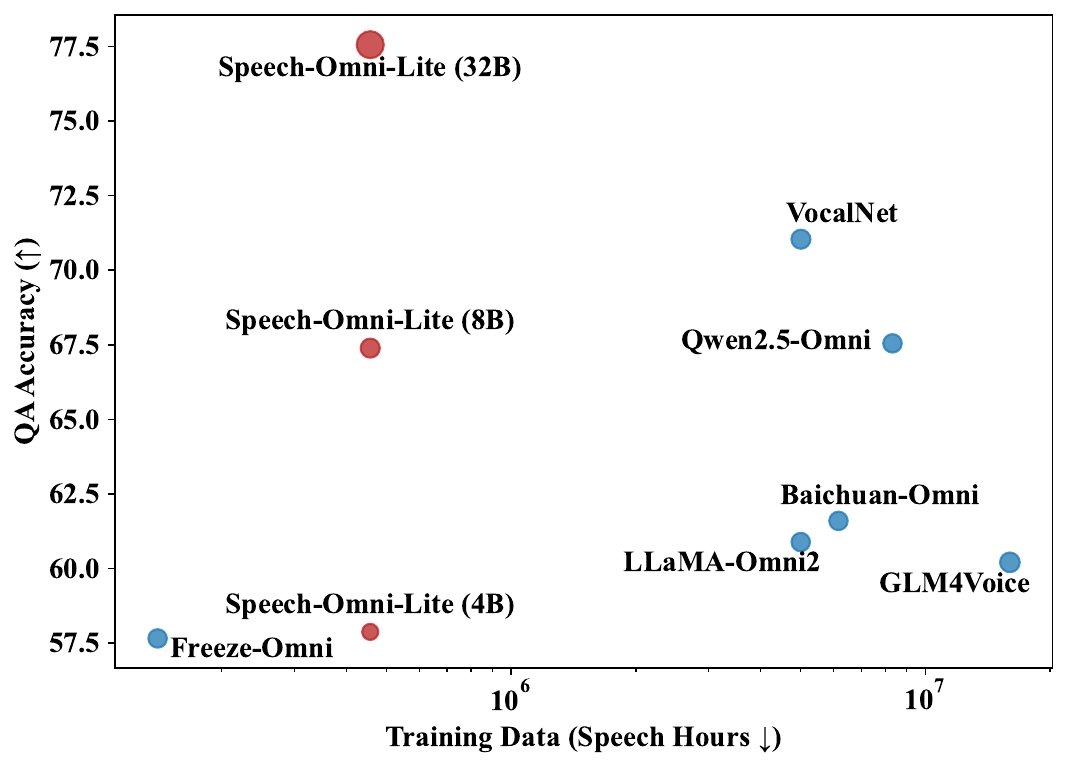}
    \caption{Comparison of training cost and question answering accuracy across different models.
Speech hours are calculated by aggregating all speech data used to endow models with speech understanding and generation capabilities. Ours achieves competitive performance with approximately one-tenth of the training cost.}
    \label{fig:placeholder}
\end{figure}

\begin{figure*}[!t]
  \vskip 0.2in
  \begin{center}
    \centerline{\includegraphics[width=0.98\linewidth]{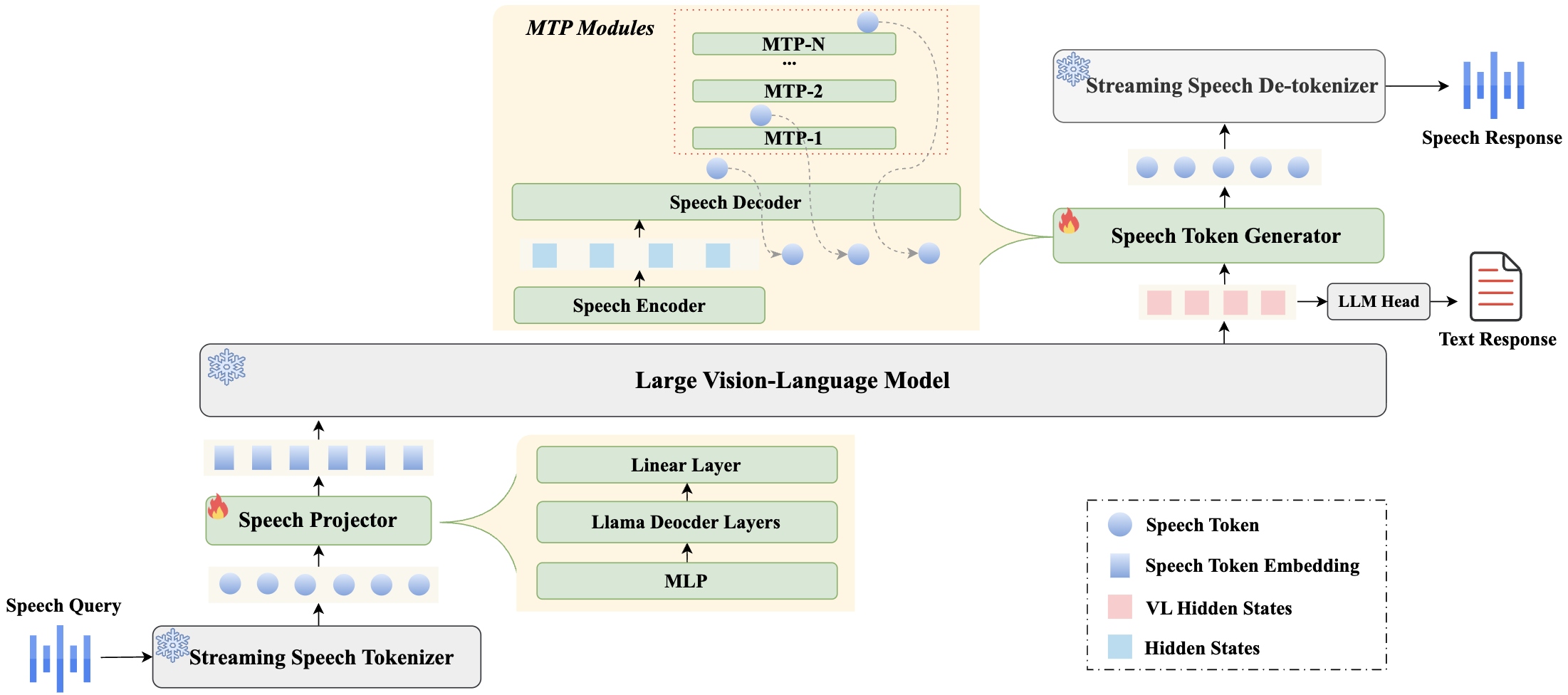}}
    \caption{The architecture of \ours. It comprises a pre-trained discrete speech tokenizer, a \textit{trainable \textbf{speech projector}}, a pre-trained large VL model, a \textit{trainable \textbf{speech token generator}}, and a pre-trained speech de-tokenizer.}
    \label{fig:speechomni}
  \end{center}
\end{figure*}

A common strategy to extend a model's modalities is to build upon a text-based large language model (LLM), attach modality-specific components, and then fine-tune the whole model for cross-modality alignment. 
This paradigm has been widely adopted in the extension of language models to the vision domain, giving rise to vision–language (VL) models such as LLaVA~\cite{liu2023visual-llava} and Qwen2-VL~\cite{wang2024qwen2vl}, where visual encoders and lightweight adaptors are integrated with strong language backbones to endow them with visual perception and reasoning capabilities. 


Beyond visual modality, \textbf{speech} serves as a primary natural medium for human–machine interaction, driving researchers to integrate it into LLMs to equip these models with the ability to ``listen'' (speech understanding) and ``speak'' (speech generation)~\cite{llama-omni2, wang-etal-2025-vocalnet}.
Existing approaches can be broadly grouped into two families. \textbf{(i) Token-in-token-out} methods discretize speech into token sequences analogous to text tokens and train a unified speech-language model that consumes and generates discrete tokens~\cite{zhang2025mimo, zeng2024glm4}. 
This provides a clean interface to the LLM but typically requires large-scale speech-text alignment data and extensive training.
\textbf{(ii) Embedding-in-token-out} methods encode speech into continuous representations and use an adaptor to map them into the LLM input space; then a speech token decoder produces speech tokens from the LLM hidden states~\cite{chen2025minmo, Qwen3-omni}. 
While effective, these methods commonly require at least partial fine-tuning of the LLM for alignment, which may significantly alter (or even degrade) the backbone performance, risking catastrophic forgetting~\cite{2025Catastrophic}.
Moreover, the learned speech adaptor and generator are often tightly coupled with the specific fine-tuned backbone, making them difficult to transfer to new backbones without retraining.

In this work, we introduce \textbf{\ours}, a cost-efficient framework to extend a pre-trained VL backbone with speech understanding and generation capabilities \textbf{while keeping the VL backbone fully frozen}. 
Our approach (Figure~\ref{fig:speechomni}) first tokenizes input speech into discrete units using a pre-trained discrete speech tokenizer. A lightweight \emph{speech projector} then maps these discrete tokens into the VL input embedding space, enabling the frozen VL backbone to process speech inputs. On the output side, a \emph{speech token generator} converts the VL hidden states into discrete speech tokens, which are subsequently transformed into waveforms by a pre-trained speech de-tokenizer. By confining speech-specific learning to lightweight and trainable modules atop a frozen VL backbone, \ours preserves the original VL capabilities while enabling speech interaction.

Beyond the model design challenge, data scarcity poses a critical obstacle to effective speech generation training.
Current speech-to-speech dialogue systems are typically trained on spoken question–answering (QA) corpora, which are expensive to collect at scale. 
To alleviate this dependency, \ours leverages \textbf{Question–Text Answer–Text–Speech (QTATS)} data, constructed from abundant existing ASR corpora via LLM-driven \textbf{reverse question generation}. This pipeline obviates the need for dedicated spoken QA recordings or speech synthesis.
To enable QTATS-based training of the speech token generator, we further introduce an auxiliary \emph{text projector} to map text inputs to the VL backbone's input space, enabling text-conditioned supervision for effective speech token generation. 

Key contributions of \ours{} are threefold:

$\bullet$ \textbf{Speech modality extension without catastrophic forgetting:} We present a framework that endows a pre-trained VL backbone with speech understanding and generation capabilities while keeping all backbone parameters frozen, thus preserving its native vision–language competence.

$\bullet$  \textbf{Lightweight and transferable speech modules:} We design compact and trainable speech modules (a speech projector and a speech token generator), along with a pre-trained speech tokenizer and a de-tokenizer. This modular design drastically reduces training overhead and enables seamless transfer across diverse VL backbones.

$\bullet$  \textbf{A novel low-cost data construction strategy for speech generation training:} 
To the best of our knowledge, we are the first to 
convert ASR speech–text pairs into spoken QA data via reverse question generation. Combined with a text-conditioned training procedure, QTATS eliminates the reliance on costly large-scale spoken QA datasets and supports effective training of the speech token generator.

\section{Related Work}
\subsection{Toward True Omni-Modal Models: Unified Support for Text, Vision, and Speech}

While omni-models are supposed to unify multiple modalities within a single foundation model, existing models vary substantially in their modality coverage.
Some ``omni''  models focus primarily on vision-language integration without native speech support, while others extend LLMs with only speech interaction capabilities. For instance, OmniVL~\cite{wang2022omnivl} is an early foundation model designed for joint image-text and video-text understanding and generation, yet it lacks native support for the speech modality. Conversely, LLaMA-Omni~\cite{fang2024llama, llama-omni2} focus on seamless speech-to-speech interaction by augmenting LLMs with speech encoders and decoders, but they do not incorporate visual perception. Although these models advance unified processing within their respective domains, they fall short of being "fully omni-modal" as they do not simultaneously support text, speech, and visual modalities in both input and output.

More recent systems move closer to fully omni-modality by jointly supporting speech, text, and vision (images and video) within a unified architecture. Baichuan-Omni-1.5~\cite{Baichuan} introduces an end-to-end multimodal framework that supports both speech understanding and speech generation, alongside visual perception and reasoning. EMOVA~\cite{chen2025emova} explores unified multimodal interaction through integrated speech, vision, and language generation within a single model architecture. Notably, EMOVA introduces end-to-end multimodal alignment with particular attention to emotional and tonal expressiveness in speech output, addressing the challenge of natural conversational interaction. Similarly, Qwen3-Omni~\cite{Qwen3-omni} extends the Qwen model family to achieve comprehensive coverage across text, speech, images, and video modalities, with full support for both input and output across all modalities without performance degradation.
Collectively, these models go beyond multimodality-in text-out pipelines and are designed for fully multimodal interaction. 

In this landscape, \ours further advances fully omni-modality by equipping strong vision-language backbones with speech understanding and generation capabilities, enabling unified multimodal reasoning and interaction while preserving native vision-language performance.



\subsection{Strategies for Speech Modality Integration}

Approaches to speech modality integration can be broadly divided into \emph{pipeline-based} and \emph{end-to-end} methods.

The simplest deployment for speech modality integration is a cascade pipeline consisting of automatic speech recognition (ASR), an LLM, and text-to-speech (TTS)~\cite{huang2024audiogpt}. This design has two fundamental limitations. First, it uses text as an intermediate representation, so the LLM only receives semantic content extracted by ASR, discarding potentially useful paralinguistic cues and fine-grained acoustic information~\cite{ward2006non, kidd2011toddlers, zhao2025moss}. Second, it suffers from high latency and error accumulation, where ASR errors propagate and are amplified in downstream generation~\cite{liu2025x}.

The end-to-end methods are categorized into two paradigms. The first paradigm represents speech using discrete tokens and incorporates them directly into the LLM vocabulary, enabling unified next-token prediction for both text and speech sequences~\cite{zhang2023speechgpt,zeng2024glm4}. This framework is conceptually simple but demands large-scale speech tokenization, elaborate vocabulary design, and extensive joint training for high-quality speech generation.

The second paradigm adopts a modular \textbf{Thinker-Talker} architecture~\cite{wang-etal-2025-vocalnet,Qwen2.5-omni,llama-omni2,wangfreeze}, which decouples speech-to-text understanding from text-to-speech generation. In this framework, a \emph{Thinker} (typically an LLM) processes speech inputs, encoded by a speech encoder and aligned to the textual space via an adaptor, for semantic understanding and reasoning. A \emph{Talker} module then generates discrete speech representations from the Thinker’s outputs, which are subsequently converted into waveforms by a token-to-speech decoder~\cite{du2024cosyvoice,ren2024fewer-codec}. This decoupling optimizes each component for its specific task, boosting scalability and generation quality.


\ours follows the Thinker-Talker paradigm to extend pre-trained vision-language backbones with dedicated speech understanding and generation modules.  By integrating speech through lightweight adaptors rather than retraining the entire foundation model, it achieves efficient speech modality expansion at ultra-low training cost.




\section{Methods}

\ours is a speech-to-speech dialogue model that augments a pre-trained VL backbone with speech input and output, while maintaining the backbone's capabilities intact. Figure~\ref{fig:speechomni} illustrates the architecture, which comprises a pre-trained discrete speech tokenizer, a \textit{trainable \textbf{speech projector}}, a pre-trained large VL model, a \textit{trainable \textbf{speech token generator}}, and a pre-trained speech de-tokenizer. The speech tokenizer, VL backbone, and de-tokenizer are kept fully frozen throughout training, while only the lightweight speech projector and speech token generator are optimized.

\ours acquires \textbf{speech understanding} by projecting discrete speech tokens into the VL backbone's input embedding space via a speech projector, enabling speech-conditioned text generation.
On the other hand, it acquires \textbf{speech generation} by converting the VL backbone's output hidden states into discrete speech tokens using a speech token generator. At inference time, these modules, together with the speech tokenizer and de-tokenizer, support end-to-end speech-to-speech interaction by interpreting speech input and synthesizing speech responses that correspond to the text produced by the VL backbone. The following subsections elaborate on the model components, the training procedure, and the data construction strategy.

\subsection{Speech input modeling for text generation}

\subsubsection{Streaming discrete speech tokenizer}
To support discrete speech-token inputs and low-latency responses, \ours employs a streaming discrete speech tokenizer that converts incoming speech into tokens chunk by chunk. The tokenizer operates at a token rate of 12.5~Hz and comprises a HuBERT LARGE encoder~\cite{hsu2021hubert}, convolutional downsampling layers, and a Finite Scalar Quantization (FSQ) module~\cite{mentzer2023finite}. The HuBERT LARGE encoder is pre-trained on 0.45~million hours of speech. Since HuBERT produces frame-level representations at 50~Hz, two convolutional downsampling layers are introduced to reduce the frame rate to 12.5~Hz. The FSQ module then maps continuous representations to discrete speech tokens.


To enable streaming training and inference, several components are refactored to operate causally on audio chunks, including: (i) convolution-based modules (the 7-layer convolutional feature extractor in HuBERT and the two downsampling convolution layers), and (ii) Transformer layers, where causal attention and chunk-wise positional embeddings are adopted. During training, FSQ code vectors are fed to a 4-layer convolutional network, and the outputs are optimized with phoneme labels using a Connectionist Temporal Classification (CTC) loss~\cite{graves2006connectionist-ctc}. This convolutional stack serves as an \emph{auxiliary} prediction head used only for tokenizer training; it is discarded after training and is not used when extracting discrete speech tokens. The training procedure follows prior discrete tokenization methods~\cite{tao2024toneunit, chen2025emova, zhang2026dsa}. 
The architectural details are presented in the Appendix~\ref{appendix:speech-tokenizer}.

\subsubsection{Speech projector}

The speech projector maps speech-token embeddings into the input embedding space of the backbone VL model. It consists of a Multi-Layer Perceptron (MLP), several LLaMA decoder layers, and a final linear projection, as shown in Figure~\ref{fig:speechomni}. The MLP lifts the speech token embedding dimension to the Transformer hidden size, and the final linear layer maps the hidden states to match the dimension of the VL input embedding. The projector is connected to the VL backbone to support streaming speech inputs.

The speech projector is trained by a two-stage scheme. First, it is trained on ASR speech–text pairs to align speech inputs with text outputs. Second, it is further trained on speech-question–text-answer data to enable speech question answering. Throughout the training process, only the speech projector parameters are updated while the discrete speech tokenizer and the VL backbone remain fully frozen.

\begin{figure}[!t]
  \vskip 0.2in
  \begin{center}
    \centerline{\includegraphics[width=\columnwidth]{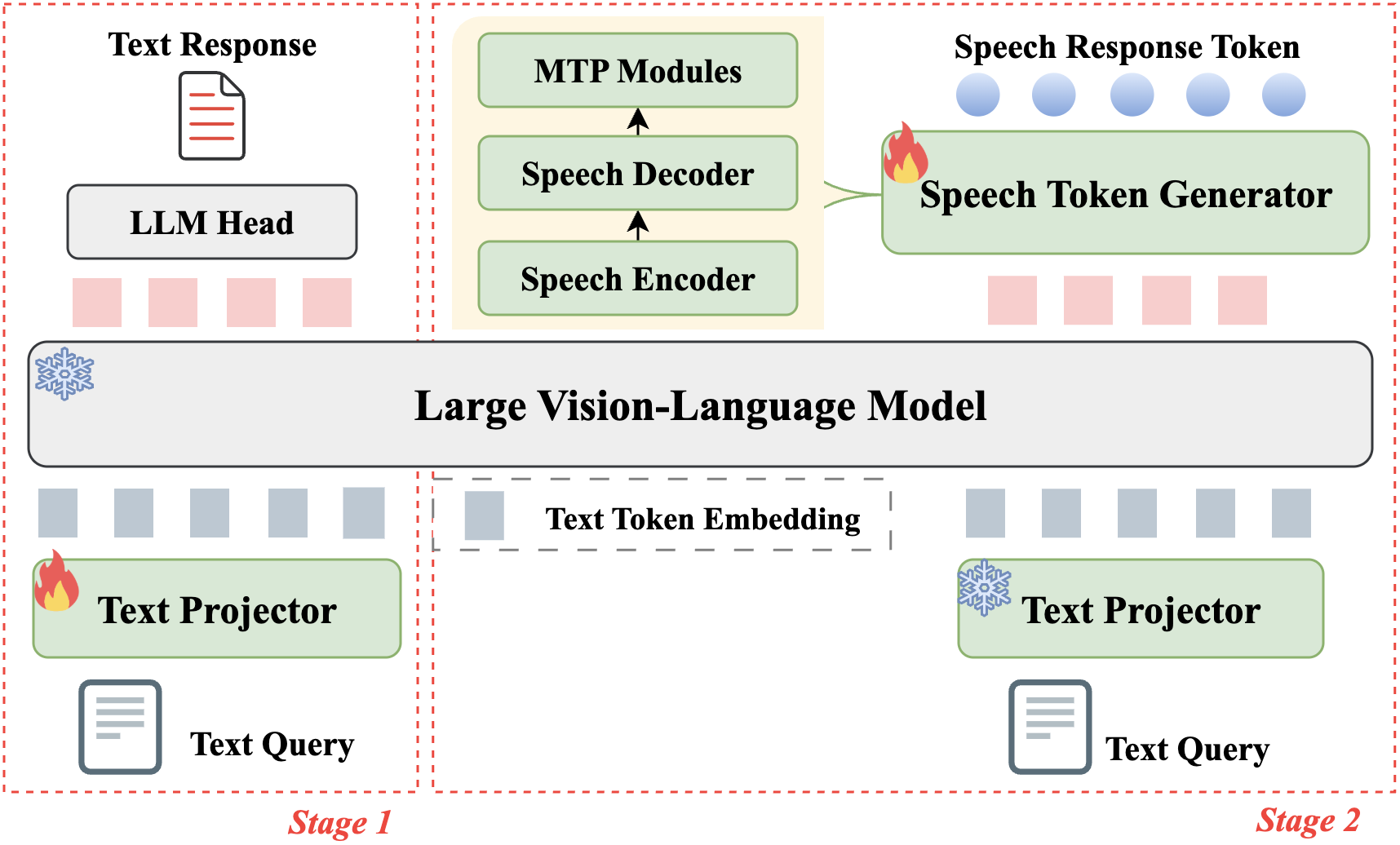}}
    \caption{The training strategy of speech token generator. In the first stage, the text projector is trained with VL backbone frozen. In the second stage, the speech token generator is trained with other modules frozen. Both stages leverage the QTATS data.}
    \label{fig:speechtokgen}
  \end{center}
\end{figure}

\subsection{Speech output modeling for speech synthesis}

\subsubsection{Speech token generator}
\label{subsubsec:tokgen}

To generate discrete speech tokens from the output hidden states of the VL backbone, a speech token generator is implemented using an encoder-decoder architecture following prior designs in VocalNet~\cite{wang-etal-2025-vocalnet} and Freeze-Omni~\cite{wangfreeze}. Multi-token prediction (MTP)~\cite{wang-etal-2025-vocalnet} is further incorporated to accelerate decoding. As illustrated in Figure~\ref{fig:speechtokgen}, the VL hidden states are first fed into a linear projection layer and then processed by an encoder. Conditioned on the encoder outputs, a decoder predicts the next speech token, after which several lightweight MTP heads predict a short span of subsequent tokens. This decoder-MTP procedure is repeated autoregressively until the complete token sequence is generated (e.g., upon emitting an end-of-sequence token or reaching a predefined length).

\textbf{Question-Text Answer-Text-Speech (QTATS) data:} The speech token generator is trained to map VL hidden states to speech tokens. A straightforward supervision strategy uses ASR corpora, treating speech as input and the corresponding discrete speech tokens as targets. In our experiments, however, a generator trained solely under ASR-style conditioning generalizes poorly to QA settings: it reconstructs tokens reliably for ASR utterances, but fails when conditioned on VL hidden states produced during QA. 
A plausible explanation is a \emph{conditioning mismatch}: VL hidden states induced by ASR-style conditioning differ from those induced by QA-style conditioning, even when the resulting text content is identical or highly similar. Meanwhile, spoken QA corpora are limited, and synthesizing spoken QA at scale via TTS can be costly. To obtain scalable supervision while avoiding dedicated spoken QA collection, QTATS is constructed from readily available ASR speech-text pairs. 
Concretely, for each ASR pair $(x^{\text{sph}}, y^{\text{txt}})$, the transcript $y^{\text{txt}}$ is treated as the \emph{answer text}. An LLM is prompted to generate a corresponding \emph{question text} conditioned on this answer. The original speech $x^{\text{sph}}$ is retained as the \emph{answer speech}, yielding triplets $(q^{\text{txt}}, a^{\text{txt}}, a^{\text{sph}})$ used to train the speech token generator under QA-style conditioning.

\textbf{Auxiliary text projector:} To make use of QTATS effectively, an auxiliary \emph{text projector} is introduced, mirroring the architecture of the speech projector but operating on text inputs. The text is first tokenized using the VL backbone's text tokenizer and then mapped to embeddings through the backbone’s input embedding matrix. These embeddings are fed into the text projector, which interfaces with the frozen VL backbone and enables QA-style conditioning without requiring spoken questions.

\begin{table}[!b]
\caption{Statistics of ASR datasets}
\label{tab:asrdata}
\begin{center}
\begin{tabular}{ccr}
\hline\hline
Language                 & Dataset        & Duration (h) \\ \hline\hline
\multirow{3}{*}{Chinese} & WenetSpeech    & 500          \\
                         & MagicData-RAMC & 128          \\ 
                         & \textit{In-house data} & 1,500         \\ \hline
\multirow{4}{*}{English} & LibriSpeech    & 500          \\
                         & GigaSpeech     & 500          \\
                         & LibriHeavy     & 500          \\
                         & CommonVoice    & 500          \\ \hline
Total                    &                & 4,128         \\ \hline\hline
\end{tabular}%
\end{center}
\end{table}

\textbf{Training strategy:} Figure~\ref{fig:speechtokgen} illustrates the training process, which is carried out in two distinct stages. First, the text projector is trained (with the VL backbone kept frozen) to support processing the QTATS question text so that the backbone produces hidden states appropriate for QA-style generation. Second, the speech token generator is attached and trained on QTATS to map the resulting VL hidden states to the target speech token sequence corresponding to the answer speech, while keeping both the VL backbone and the text projector frozen. After training, the text projector is no longer needed and can be removed.

\subsubsection{Speech De-Tokenizer}
Speech de-tokenization synthesizes waveform audio from the discrete speech tokens predicted by the speech token generator. he de‑tokenizer follows the overall design of F5‑TTS~\cite{chen2025f5-tts}, but replaces its DiT module with a cross‑attention‑augmented variant, referred to as CA‑DiT. In particular, the noisy and masked acoustic representations are concatenated and passed to the CA-DiT blocks as the primary input, while the speech tokens serve as the conditioning signal through cross-attention (as key-value pairs). To mitigate the frame-rate mismatch between acoustic features and token sequences (e.g., Mel features at $\sim$94~Hz versus tokens at $\sim$12.5~Hz), the token sequence is temporally upsampled to align more closely with the Mel frame rate before conditioning. Architectural details are presented in the Appendix~\ref{appendix:de-tokenizer}.









\section{Experimental Setup}

\subsection{Datasets}

In this work, approximately 4,000 hours of speech-text paired ASR data, covering both Chinese and English, are used to train the speech tokenizer and the speech projector. This corpus is sampled from 6 open-source datasets (2~in Chinese and 4~in English); we additionally supplement 1,500 hours of in-house Chinese data to ensure roughly balanced Chinese and English data volumes in the final corpus, with dataset-level statistics reported in Table~\ref{tab:asrdata}. For speech tokenizer training, CTC~\cite{graves2006connectionist-ctc} supervision is derived by converting each transcript into phoneme sequences using a G2P model. For Chinese, the resulting phonemes are used directly as CTC targets. For English, in-word BPE segments are used as CTC targets by applying BPE to the phoneme sequence within each word. For speech projector training, the output supervision of the VL backbone is the transcript corresponding to the input speech.


To construct the QTATS data described in Section~\ref{subsubsec:tokgen} from the ASR corpus, Qwen3-30B-A3B-Instruct-2507\footnote{{https://huggingface.co/Qwen/Qwen3-30B-A3B-Instruct-2507}} is used to generate a \emph{question} for each speech-text pair whose transcript can serve as an \emph{answer}. This process yields approximately 1.45 million QTATS question-answer pairs, which are used to train the auxiliary text projector and the speech token generator.
Additionally, to further enhance the spoken QA capability of \ours, we use VoiceAssistant-430K dataset, which is a cleaned variant of VoiceAssistant-400K with overlong responses removed, containing 430K speech-query-text-response pairs.


The speech de-tokenizer is trained on around 6k speech hours, including LibriTTS~\cite{zen2019libritts}, WenetSpeech4TTS~\cite{ma2024wenetspeech4tts}, and in-house data. Discrete speech tokens are extracted from audio using the trained speech tokenizer under a chunk-wise inference setting, and serve as conditioning inputs for de-tokenizer training.

\subsection{Model configuration}

\textbf{Streaming discrete speech tokenizer:} HuBERT LARGE is used as the frontend speech encoder, followed by two convolutional layers with an overall $2\times$ temporal downsampling, and an FSQ module with quantization levels $[8,8,8,8,8]$. A four-layer convolutional \emph{phoneme decoder} (with stride equals $1$) is introduced as an auxiliary head to map FSQ code vectors to representations suitable for computing the CTC loss. The speech chunk size is set to 640~ms, and each chunk attends to the previous 30 chunks during streaming training. The same configuration is used for speech token extraction. Unless otherwise noted, the phoneme decoder is used only during tokenizer training and is discarded for token extraction. The speech tokenizer (excluding the phoneme decoder) contains approximately 337M parameters and outputs tokens at 12.5~Hz.

\textbf{Speech projector:} The speech projector comprises (i) a 2-layer MLP with GELU activation function that maps each 5-dimensional FSQ code vector to a 2048-dimensional representation; (ii) 6 Llama-style decoder layers with hidden size 2048, 32 attention heads, and an 8192-dimensional FFN; and (iii) a final linear layer that projects the decoder outputs to the VL input embedding dimension (i.e., to match the backbone's input embedding size).
In total, the speech projector contains approximately 415M  parameters, including those in the final projection layer.

\textbf{Backbone VL:} In this work, Qwen3‑VL‑8B‑Instruct\footnote{{https://huggingface.co/Qwen/Qwen3-VL-8B-Instruct}} is used as the backbone. Since the backbone was fully frozen throughout the training process, \ours is inherently backbone‑agnostic: its components can be integrated with various pretrained backbones without retraining or modifying the backbone itself (see Section~\ref{subsec:trans}), facilitating scalable deployment across different model families. 

\textbf{Speech token generator:} The speech token generator configuration follows the framework established in VocalNet. It adopts an encoder-decoder design with an 8-layer speech encoder and a 4-layer speech decoder. Both components are instantiated using Llama-style decoder blocks with 32 attention heads. The hidden size and Feed-Forward Network (FFN) dimension are set to 4096 and 16384, respectively. The model incorporates five MTP modules, each implemented as a single Llama‑style decoder layer followed by a linear output head. During inference, the number of MTP modules is reduced to two, as this configuration yields the best performance according to experimental results.

\textbf{Speech de-tokenizer:} CA-F5-TTS is based on F5-TTS, using a CA-DiT backbone with 22 layers, 16 attention heads, and embedding/FFN dimensions of 1024/2048, respectively. To reduce the rate mismatch between speech tokens and acoustic frames, the speech token sequence is upsampled by a factor of 6, resulting in a token rate of 75~Hz, which more closely matches the Mel-spectrogram frame rate ($\sim$94~Hz). In our experiments, audio is synthesized in a chunk‑by‑chunk manner during inference, using a chunk size of 25 tokens, corresponding to 2-second audio.

\begin{table*}[tb]
\caption{Results on \textbf{ASR} and \textbf{QA} tasks.
AS-1 is short for AISHELL-1; WenetS. for WenetSpeech; LibriS. for LibriSpeech.  ``/" denotes that the model does not possess the ASR ability. ``*" denotes that only the final linear projection layer of the speech projector is trained.
Results for GLM-4-Voice are taken from the original paper, as we were unable to reproduce its ASR function. \textbf{Bold} denotes the best result within each subgroup, and \underline{underline} denotes the second best.}
\label{tab:s2tres}
\begin{center}
\setlength{\tabcolsep}{4pt}
\resizebox{\textwidth}{!}{%
\begin{tabular}{c|rrrrr|cccccccc}
\hline
\multirow{3}{*}{Model} & 
\multicolumn{5}{c|}{ASR (\%CER $\downarrow$; \%WER $\downarrow$)} & \multicolumn{8}{c}{Spoken QA (\%Accuracy $\uparrow$)} \\

 & AS-1 & \multicolumn{2}{r}{WenetS. (test)} & \multicolumn{2}{r|}{LibriS. (test)} & \multicolumn{2}{c}{Llama Questions} & \multicolumn{2}{c}{Web Questions} & \multicolumn{2}{c}{TriviaQA} & \multicolumn{2}{c}{AlpacaEval} \\
 
 & test & net & meeting & clean & other & S-\textgreater{}T & S-\textgreater{}S & S-\textgreater{}T & S-\textgreater{}S & S-\textgreater{}T & S-\textgreater{}S & S-\textgreater{}T & S-\textgreater{}S \\ \hline
 
GLM-4-Voice (9B) 
& 2.46 & - & - & 2.82 & 7.66 & 76.67 & 63.67 & 49.50 & 46.90 & 49.50 & 44.84 & 65.15 & 58.59 \\

VocalNet (8B) 
& \multicolumn{5}{c|}{/} & 83.33 & \textbf{78.60} & \underline{62.90} & \underline{59.80} & \underline{62.16} & \textbf{57.30} & 75.76 & \textbf{72.22} \\

LLaMA-Omni2 (7B) 
& \multicolumn{5}{c|}{/} & 79.33 & 72.00 & 51.40 & 49.20 & 48.70 & 44.84 & 64.14 & 63.64 \\

Freeze-Omni (7B) 
& \multicolumn{5}{c|}{/} & 78.67 & 58.67 & 54.10 & 29.80 & 51.15 & 32.36 & 46.73 & 28.79 \\

Baichuan-Omni-1.5 (7B) 
& \underline{2.37} & \underline{8.70} & \underline{9.86} & \underline{2.54} & \underline{5.63} & 77.67 & 65.67 & 49.90 & 32.39 & 54.70 & 47.28 & 64.14 & 13.92 \\

Qwen2.5-Omni-7B 
& \textbf{1.12} & \textbf{6.20} & \textbf{7.80} & \textbf{2.31} & \textbf{4.19} & 79.67 & \underline{77.33} & 61.60 & \textbf{60.10} & 57.20 & \underline{56.30} & 71.72 & \underline{67.17} \\ \hline

\ours (8B) 
& 5.77 & 16.54 & 18.32 & 4.39 & 9.22 & \underline{85.00} & 73.33 & 51.50 & 46.20 & 56.80 & 50.40 & \underline{76.26} & 65.66 \\ \hline


\ours (4B)* 
& 9.01 & 19.92 & 20.74 & 5.76 & 10.81 & 78.33 & - & 43.70 & - & 41.30 & - & 68.18 & - \\

\ours (32B)* 
& 5.46 & 16.50 & 18.18 & 3.58 & 7.75 & \textbf{86.00} & - & \textbf{66.60} & - & \textbf{71.80} & - & \textbf{85.86} & - \\ \hline

\end{tabular}%
}
\end{center}
\end{table*}

\subsection{Training and evaluation setup}

\textbf{Training:} \ours is trained in two independent phases: (i) speech projector training and (ii) speech token generator training. For speech projector training, optimization is conducted in two stages. In Stage~1, the full ASR corpus is used to endow the model with ASR capability. In Stage~2, $10\%$ of the ASR corpus plus the full spoken QA corpus are used to adapt the model to speech question answering. For speech token generator training, the full QTATS corpus is used in two steps: the auxiliary text projector is trained first, and then the speech token generator is trained while keeping the text projector frozen. Across both training phases, the learning rate is set to $2\times10^{-4}$ with a cosine annealing schedule and a warmup ratio of $0.03$. Since the two phases are independent, the speech projector and the speech token generator can be trained separately or in parallel using the same frozen VL backbone, and attached to the input and output sides of the backbone after training.

\textbf{Speech understanding evaluation (S$\rightarrow$T):} Speech understanding is evaluated on ASR and speech-to-text QA tasks, where speech audio is the input and text is the output. For ASR, evaluation is performed on three Chinese test-sets, including \textit{AISHELL-1 test}, \textit{WenetSpeech test-net}, \textit{WenetSpeech test-meeting}, and two English test-sets, including \textit{LibriSpeech test-clean}, and \textit{LibriSpeech test-other}. ASR performance is reported using character error rate (CER) for Chinese and word error rate (WER) for English. For QA, evaluation is conducted on \textit{LLaMA Questions}, \textit{Web Questions}, \textit{TriviaQA}, and \textit{AlpacaEval}, with accuracy reported as the percentage of correct answers judged by GPT-5.

\textbf{Speech generation evaluation (S$\rightarrow$S):} Speech generation is evaluated on the same four QA sets in the speech-to-speech setting. Two aspects are assessed: (i) content consistency between the model's generated text and the synthesized speech, and (ii) perceptual audio quality. The synthesized speech is transcribed using Whisper-large-v3.\footnote{https://huggingface.co/openai/whisper-large-v3} Content consistency is measured by (1) GPT-5 scoring of the transcribed response and (2) the word error rate (WER) between the model-generated text and the transcription. Audio quality is measured using UTMOS~\cite{saeki2022utmos}.


\section{Experimental Results}

\subsection{Results on speech context to text generation}

Table~\ref{tab:s2tres} reports ASR and speech-to-text QA results on English and Chinese evaluation sets for \ours and representative speech LMs and omni models. For a fair comparison in streaming settings, the streaming chunk size is fixed to 640~ms, matching the \ours configuration for all models that support streaming speech input. An empirical observation is that some prior speech LMs and omni models primarily target QA and lack ASR capability (i.e., VocalNet, LLaMA-Omni2, and Freeze-Omni\footnote{While the paper reported the ASR results, the released model cannot perform the ASR task, according to the author's response in https://github.com/VITA-MLLM/Freeze-Omni/issues/10 .}), which can limit coverage of speech understanding tasks. Owing to the comparatively smaller amount of ASR data used to train the speech tokenizer and speech projector, \ours underperforms large-scale omni models trained with substantially more speech data (i.e., Baichuan-Omni-1.5, Qwen2.5-Omni) on ASR benchmarks. Nevertheless, \ours achieves strong QA performance, particularly on \textit{LLaMA Questions} and \textit{AlpacaEval}. Overall, these results suggest that, with only a trainable speech projector trained on limited ASR and QA data, \ours attains both ASR functionality and competitive spoken QA performance relative to large-scale omni models trained with millions of hours of speech.

\begin{table*}[htb]
\caption{Results on \textbf{content consistency} (WER $\downarrow$) between speech response and text response and \textbf{acoustic performance} (UTMOS $\uparrow$). \textbf{Bold} denotes the best result within each subgroup, and \underline{underline} denotes the second best.}
\label{tab:s2sres}
\begin{center}
\resizebox{\linewidth}{!}{%
\begin{tabular}{l|rcrcrcrc|rc}
\hline
\multirow{2}{*}{Model} & \multicolumn{2}{c}{Llama Questions} & \multicolumn{2}{c}{Web Questions} & \multicolumn{2}{c}{TriviaQA} & \multicolumn{2}{c|}{AlpacaEval} & \multicolumn{2}{c}{Average} \\
 & WER & UTMOS & WER & UTMOS & WER & UTMOS & WER & UTMOS & WER & UTMOS  \\ \hline
 
GLM-4-Voice (9B) & 41.00 & 4.15 & 10.16 & 4.11 & 13.14 & 4.15 & 29.44 & 3.86 & 23.44 & 4.07 \\

VocalNet(8B) & \underline{2.79} & \textbf{4.47} & \underline{5.01} & \textbf{4.49} & \underline{4.86} & \textbf{4.47} & 7.43 & \textbf{4.49} & 5.02 & \textbf{4.48} \\

LLaMA-Omni2 (7B) & 3.77 & \underline{4.44} & 5.75 & \underline{4.43} & 6.02 & \underline{4.44} & \textbf{3.88} & \underline{4.43} & \underline{4.86} & \underline{4.44} \\

Freeze-Omni (7B) & 24.34 & 4.36 & 47.31 & 4.23 & 43.61 & 4.26 & 49.47& 4.14 & 41.18 & 4.25 \\

Baichuan-Omni-1.5 (7B) & 31.64 & 4.31 & 45.58 & 4.32 & 29.51 & 4.35 & 86.53 & 4.30 & 48.32 & 4.32 \\

Qwen2.5-Omni-7B & \textbf{0.99} & 4.30 & \textbf{1.68} & 4.32 & \textbf{1.47} & 4.29 & \underline{5.54} & 4.20 & \textbf{2.42} & 4.28 \\ \hline
\ours (8B) & 12.49 & 4.12 & 22.55 & 3.96 & 15.04 & 4.08 & 23.77 & 3.73 & 18.46 & 3.97 \\ \hline
\end{tabular}%
}
\end{center}
\end{table*}

\subsection{Results on speech context to speech generation}

The evaluation results for content consistency between text and generated speech, as well as acoustic quality, are summarized in Table~\ref{tab:s2tres} and Table~\ref{tab:s2sres}. Although \ours attains mid-range WER and UTMOS scores compared to representative streaming speechLMs and omni-models, it achieves competitive speech-to-speech QA performance. Importantly, this level of performance is obtained under a markedly more economical supervision setting than existing systems. The speech token generator is trained entirely on manufactured QA-style supervision derived from readily accessible ASR corpora, in contrast to prior work that depends on large volumes of curated or synthetic spoken QA pairs. Because such manufactured supervision lacks genuine QA reasoning structure, a training–inference mismatch in the VL hidden states conditioning speech generation is unavoidable. Despite this limitation, the empirical results show that the proposed strategy remains highly effective—robust speech generation behavior emerges even without spoken QA corpora. This outcome demonstrates that high-quality speech generation for the Talker structure can be obtained from a simple, scalable, and low-cost supervision pipeline built solely from ASR resources.

\subsection{Transferability of speech projector}
\label{subsec:trans}

To examine whether a speech projector trained with the Qwen3-VL-8B-Instruct backbone can transfer effectively to VL backbones of other sizes, 
Qwen3-VL-4B-Instruct\footnote{{https://huggingface.co/Qwen/Qwen3-VL-4B-Instruct}} and 
Qwen3-VL-32B-Instruct\footnote{{https://huggingface.co/Qwen/Qwen3-VL-32B-Instruct}} 
are selected as alternative backbones. 
When applying the well-trained speech projector to these backbones, only the final linear projection layer is updated to match the backbone’s input dimension, while all other components remain frozen. The training data and configurations follow those used in the second training stage of the speech projector. The corresponding speech understanding results are provided in Table~\ref{tab:s2tres}. Performance improves consistently as the backbone size increases, with \ours (32B) achieving the strongest speech-to-text QA results. These findings indicate that, under a frozen-backbone setting, a speech projector trained with a small VL backbone can be effectively transferred to larger backbones to endow them with speech understanding capability, substantially reducing the training cost otherwise required to scale across backbone sizes.


\subsection{Latency Analysis of the speech response pipeline}

The response latency of \ours can be decomposed into four stages: (1) the streaming speech tokenizer converts the input speech query into discrete tokens; (2) the VL backbone generates hidden states; (3) the speech token generator produces response speech tokens; and (4) the speech de‑tokenizer synthesizes the output waveform. The latency for generating a 1‑second response is summarized in Table~\ref{tab:latency} in Appendix~\ref{appendix:latency}. The speech de‑tokenizer is the dominant contributor to end‑to‑end latency, and its latency is computed under the configuration where audio is synthesized chunk‑by‑chunk from speech tokens using a chunk size of 25 tokens (corresponding to 2 seconds of audio). A key property of the system is that the speech tokenizer operates fully in streaming mode: each 640~ms speech chunk incurs a fixed 54.3~ms processing time, yielding a constant latency regardless of the input speech's duration.


\section{Discussion}

\textbf{Architectural impact on the speech projector:} The experimental results indicate that the architecture of the speech projector plays a critical role when both the VL backbone and the speech tokenizer remain frozen. Table~\ref{tab:projstruc} in Appendix~\ref{appendix:proj-struct} summarizes the ASR performance obtained in the first training stage of the speech projector. The results show that more expressive projector architectures consistently yield better performance, with Transformer‑based projectors achieving the strongest results within the \ours framework. It is also observed that simpler architectures require substantially larger learning rates to train effectively (e.g., \(10^{-3}\) or \(10^{-2}\) for MLP‑ and CNN-based projectors), whereas Transformer‑based projectors train stably with smaller learning rates (e.g., \(2\times10^{-4}\)).

\textbf{Encoder–Decoder configuration effects in the speech token generator:} The two main components of the speech token generator—the speech encoder and the speech decoder—exhibit different impacts on the quality of the generated speech tokens. The comparison results in Table~\ref{tab:genstruc} in Appendix~\ref{appendix:gen-struct} reveal two key observations. First, when the total number of Transformer layers is fixed, allocating more layers to the speech encoder leads to better performance than allocating them to the speech decoder. Second, a deeper speech encoder is more effective at bridging the representational gap between the VL backbone’s hidden states and the target speech token embeddings. These findings highlight the importance of encoder capacity in aligning VL representations with the speech token space.

\section{Conclusion}

This work presents \ours, a cost-efficient framework that augments a pretrained VL backbone with speech understanding and speech generation capabilities while preserving its original vision and language performance. To avoid reliance on costly spoken QA corpora, QTATA data are constructed from readily available ASR resources to supervise the training of the speech token generator, offering a significantly cheaper alternative to large‑scale audio synthesis. Experimental results demonstrate that, even with several thousand hours of speech data, \ours attains competitive spoken QA ability compared with mainstream speech language models and omni‑models trained on millions of hours of speech data. Furthermore, a speech projector trained with one backbone can be effectively transferred to backbones of different sizes, with speech understanding performance improving consistently as backbone capacity increases.

\section*{Impact Statement}

This paper presents \ours, a framework designed to advance the field of multimodal Machine Learning by enabling cost-efficient and portable speech interfaces for Vision-Language (VL) models. Beyond its academic contributions, our work has several potential broader impacts and societal consequences:

\begin{enumerate}
    \item \textbf{Democratization of Multimodal Research.} By utilizing lightweight, ``plug-and-play" modules and keeping the large VL backbone frozen, our approach drastically reduces the computational resources and massive datasets typically required to train omni-modal models. This helps lower the barrier to entry for smaller research groups and institutions with limited hardware budgets.
    
    \item \textbf{Resource Efficiency and Sustainability.} The proposed QTATS data construction strategy allows for effective training using approximately one-tenth of the speech data required by traditional models. This reduction in training time and data dependency translates to a smaller carbon footprint and lower energy consumption during the model development phase.

    \item \textbf{Accessibility and Human-Machine Interaction.} Speech is a primary natural medium for human interaction. By providing a portable and efficient way to add speech capabilities to existing models, this work can facilitate the development of more accessible AI interfaces for individuals with visual or motor impairments who rely on voice-based communication.

    \item \textbf{Ethical Considerations and Risk Mitigation.} While our framework enhances model capabilities, it also inherits the ethical challenges of large-scale generative models, such as the potential for generating biased or harmful content. However, because our method preserves the original backbone’s performance without retraining the entire model, it avoids ``catastrophic forgetting" and allows for the continued use of existing safety alignments within the frozen VL backbone.
\end{enumerate}




\nocite{langley00}

\bibliography{reference}
\bibliographystyle{icml2026}

\appendix
\section{Streaming Discrete Speech Tokenizer \label{appendix:speech-tokenizer}}



\begin{figure}[ht]
  \vskip 0.2in
  \begin{center}
    \centerline{\includegraphics[width=\columnwidth]{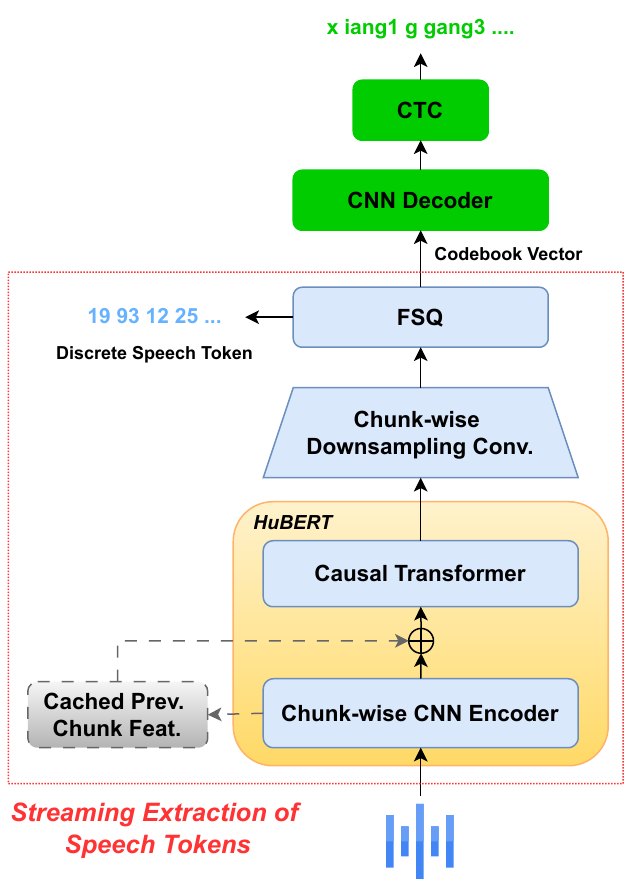}}
    \caption{Architecture of streaming discrete speech tokenizer.}
    \label{fig:sphtokenizer}
  \end{center}
\end{figure}

The training methodology for the speech tokenizer can be regarded as a task that finetunes a pretrained HuBERT LARGE encoder into a tonal phoneme recognizer. As illustrated in Figure~\ref{fig:sphtokenizer}, the raw speech waveform is first processed by the HuBERT LARGE encoder, after which the continuous representations are quantized by an FSQ module to obtain codebook vectors. These quantized embeddings are then passed to a CNN‑based phoneme decoder that predicts phoneme labels. A CTC loss is applied during training, guiding the quantized representations to encode phonetic and lexical cues from the speech signal. This design ensures that the resulting speech tokens retain the semantic information present in speech.

\section{Speech De-Tokenizer 
\label{appendix:de-tokenizer}}

\begin{figure}[ht]
  \vskip 0.2in
  \begin{center}
    \centerline{\includegraphics[width=\columnwidth]{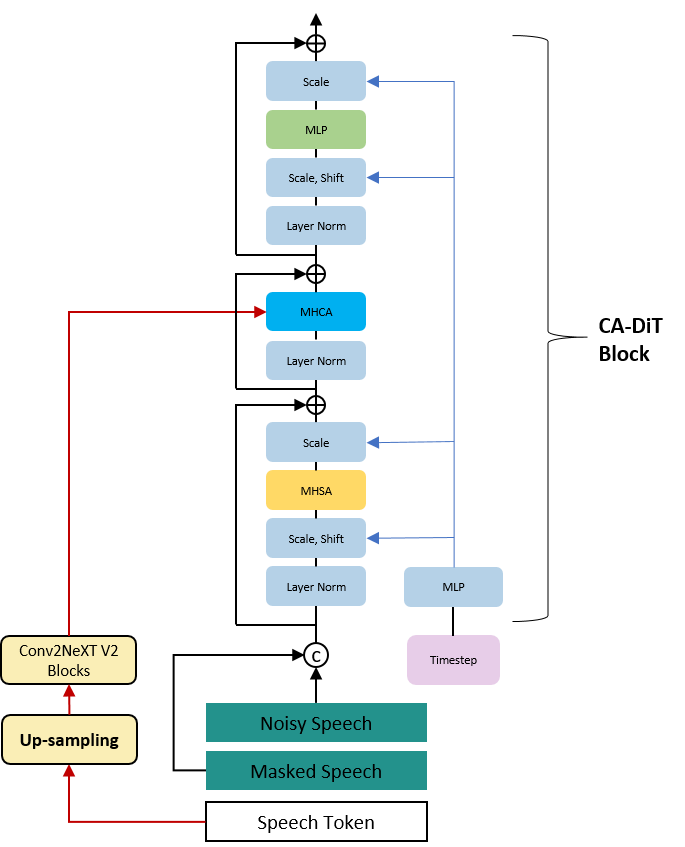}}
    \caption{Architecture of CA-DiT block of speech de-tokenizer.}
    \label{fig:sphdetoken}
  \end{center}
\end{figure}

To convert discrete speech tokens into continuous mel spectrograms, we design a dedicated speech de-tokenizer similar to F5-TTS framework while adapting to token-based conditioning. The model utilizes a CA-DiT backbone, as shown in Figure~\ref{fig:sphdetoken}, and is trained using a flow-matching paradigm. 

The CA-DiT differs from the DiT backbone of the F5-TTS in two key ways:
First, it replaces text inputs with up-sampling discrete speech tokens as the primary conditioning signal. Second, we adopt a query-key-value (QKV) conditional injection mechanism: mel spectrogram features (serving as queries) interact with discrete speech tokens (as keys and values) via cross-attention, replacing F5-TTS’s concatenation of noisy mel, clean mel, and text features. This decoupled design enhances token-acoustic alignment by explicitly modeling semantic interactions between discrete tokens and continuous mel frames.

\section{End-to-End Latency Analysis
\label{appendix:latency}}

Table~\ref{tab:latency} reports the end‑to‑end latency of generating a 1‑second speech response with \ours (8B). Latency is measured for each component in the response pipeline, including the streaming speech tokenizer, VL backbone, speech token generator, and speech de‑tokenizer.

\begin{table}[htb]
\caption{Latency breakdown (ms) for generating a 1‑second speech response using Speech‑Omni‑Lite (8B).}
\label{tab:latency}
\begin{center}
\resizebox{\linewidth}{!}{%
\begin{tabular}{ccccc}
\hline
\begin{tabular}[c]{@{}c@{}}Speech \\ Tokenizer\end{tabular} & VL & \begin{tabular}[c]{@{}c@{}}Speech \\ Token Generator\end{tabular} & \begin{tabular}[c]{@{}c@{}}Speech \\ De-tokenizer\end{tabular} & Sum \\ \hline\hline
54.32 & 123.16 & 29.33 & 1139 & 1345.79 \\ \hline
\end{tabular}%
}
\end{center}
\end{table}

\section{Detailed Analysis of Speech Projector Architectures\label{appendix:proj-struct}}



This appendix provides the experimental configurations and complete numerical results supporting the architectural comparison of the speech projector. The VL backbone (Qwen2.5‑3B\footnote{https://huggingface.co/Qwen/Qwen2.5-3B}) and the speech tokenizer remain frozen, and only the speech projector is optimized using the ASR data. The evaluated architectures include MLP‑, CNN‑, and Transformer‑based projectors with varying depths and learning‑rate settings. 

Table~\ref{tab:projstruc} reports the exact combinations of model structure, number of layers, and learning rates used in the experiments, together with WER and CER results on LibriSpeech (LS) test‑clean and AISHELL2 (AS2) test. These results provide the empirical basis for analyzing how projector expressiveness and training stability vary across architectural choices.

\section{Detailed Analysis of Speech Token Generator Configurations}
\label{appendix:gen-struct}



This appendix reports the experimental configurations and numerical results that support the layer‑allocation analysis discussed in the main text. Only the speech token generator is trained, with the VL backbone frozen. 

Table~\ref{tab:genstruc} presents the exact encoder–decoder layer configurations evaluated for Qwen3-VL‑4B and Qwen3-VL‑8B, along with their WER computed between the model‑generated text response and the transcription of the generated speech obtained using Whisper‑large‑v3. These results provide the empirical basis for analyzing how the allocation of layers between the encoder and decoder affects speech token generation quality.

\begin{table}[!tb]
\caption{ASR performance of different speech projector architectures using the Qwen2.5‑3B backbone. 
Only the speech projector is trained, while the VL backbone and speech tokenizer remain frozen. 
The projector hidden size is fixed at 2048.}
\label{tab:projstruc}
\begin{center}
\resizebox{\linewidth}{!}{%
\begin{tabular}{c|c|c|cc}
\hline
\begin{tabular}[c]{@{}c@{}}Structure of \\  Speech Projector\end{tabular} & \begin{tabular}[c]{@{}c@{}}Number of \\ Layers\end{tabular} & \begin{tabular}[c]{@{}c@{}}Learning \\ Rate\end{tabular} & \begin{tabular}[c]{@{}c@{}}WER (\%) on\\ LS test-clean\end{tabular} & \begin{tabular}[c]{@{}c@{}}CER(\%) on\\ AS2 test\end{tabular} \\ \hline\hline
\multirow{5}{*}{MLP} & 2 & 1e-2 & 19.4 & 28.6 \\
 & 3 & 1e-2 & 11.6 & 19.5 \\
 & 4 & 1e-2 & \multicolumn{2}{c}{Failed} \\
 & 4 & 1e-3 & 9.2 & 20.3 \\
 & 5 & 1e-3 & 12.3 & 20.5 \\ \hline
\multirow{3}{*}{CNN} & 4 & 1e-2 & 6.9 & 14.5 \\
 & 4 & 1e-3 & 6.6 & 14.5 \\
 & 4 & 2e-4 & 8.3 & 19.7 \\ \hline
\multirow{5}{*}{Transformer} & 2 & 1e-3 & 4.4 & 8.6 \\
 & 2 & 2e-4 & 4.3 & 8.6 \\
 & 4 & 1e-3 & \multicolumn{2}{c}{Failed} \\
 & 4 & 2e-4 & 3.9 & 7.7 \\
 & 6 & 2e-4 & 3.9 & 7.4 \\ \hline\hline
\end{tabular}%
}
\end{center}
\end{table}

\begin{table}[htb]
\caption{Comparison for different allocations of Transformer layers between the speech encoder and speech decoder in the speech token generator. 
The VL backbone is frozen, and only the token generator is trained. 
Each row corresponds to a specific encoder–decoder depth configuration evaluated in the study.
}
\label{tab:genstruc}
\begin{center}
\resizebox{\linewidth}{!}{%
\begin{tabular}{cccc}
\hline
Backbone & \begin{tabular}[c]{@{}c@{}}Speech \\ Encoder\end{tabular} & \begin{tabular}[c]{@{}c@{}}Speech \\ Decoder\end{tabular} & \begin{tabular}[c]{@{}c@{}}WER (\%) on \\ Speech Response\end{tabular} \\ \hline\hline
\multirow{5}{*}{Qwen3-VL-4B} & 2 & 4 & 17.40 \\
 & 2 & 6 & 16.40 \\
 & 4 & 4 & 14.15 \\
 & 6 & 4 & 12.00 \\
 & 8 & 4 & 12.52 \\ \hline
\multirow{3}{*}{Qwen3-VL-8B} & 6 & 4 & 14.86 \\
 & 8 & 4 & 12.49 \\
 & 10 & 4 & 17.33 \\ \hline\hline
\end{tabular}%
}
\end{center}
\end{table}

\end{document}